\documentstyle[12pt,aaspp4,flushrt]{article}

\begin{document}

\title{A New Weak Lensing Analysis of MS1224.7+2007}

\author{Philippe Fischer\footnote{Based in part on research carried out at the
MDM Observatory, operated by Columbia University, Dartmouth College, University
of Michigan and Ohio State University}}

\affil{Dept. of Astronomy, University of Michigan, Ann Arbor, MI 48109}

\slugcomment{Astronomical Journal, submitted}

\abstract

Galaxy cluster mass distributions are useful probes of $\Omega_0$ and the
nature of the dark matter. Large clusters will distort the observed shapes of
background galaxies through gravitational lensing allowing the measurement of
the cluster mass distributions. For most cases, the agreement between weak
lensing and radial velocity mass measurements of clusters is reasonably good.
There is, however, one significant exception, the $z=0.32$ cluster MS1224+2007,
which has a lensing mass substantially larger than the virial mass and also a
very high mass-to-light ratio. Since this controversial object might be an
unusually dark mass a follow-up study is definitely warranted.

In this paper we study the mass and light distributions of MS1224+2007 out to a
projected radius of 800 $h^{-1}$ kpc by measuring the gravitationally-induced
distortions of background galaxies. We detect a shear signal in the background
galaxies in the radial range $27.5\arcsec\ \le r \le 275\arcsec$ at the
5.5$\sigma$ level. The resultant mass map exhibits a peak centered on the
dominant cluster galaxy and strong evidence for substructure which is even more
strongly seen in the galaxy distribution.  Assuming all the detected shear is
due to mass at $z=0.32$ we find cluster mass-to-light ratio of M/L$_R = 640 \pm
150$ (M/L$_R)_\odot$.  The mass profile is quite flat compared to other
clusters, disagreeing with a pseudo-singular isothermal sphere at the 95\%
confidence level.
Our mass and M/L estimates are consistent with the previous weak lensing
result.  The discrepancy between the lensing and virial mass remains although
it might be partially explained by subclustering and infall perpendicular to
the line-of-site. This cluster remains a candidate dark object deficient in
baryons and as such severely tests cosmological models.


\section{Introduction}

Weak lensing distortions of background galaxies are a powerful tool for
measuring galaxy cluster mass distributions. Lensing based studies are {\it
direct} measurements and are not model-dependent as are other techniques
(X-ray, radial velocities).  Currently there are around 20 clusters which are
well-studied with lensing (e.g. \cite{fi97}), most yielding an evolution
corrected M/L$\approx300$ (M/L)$_\odot$.

There is, however, one significant exception, the z=0.32 cluster MS1224+2007,
which has M/L = 800 (M/L$)_\odot$ from the weak lensing measurements of
\cite{fa94}. Not only is the lensing-derived M/L estimate very high but the
mass estimate is approximately a factor of 2.5 higher than the virial mass
estimate (\cite{ca94}). If the lensing mass is correct then it would indicate a
large region with an anomolously low baryon fraction. Within the standard
cosmological model there is no causal mechanism which can generate primordial
fluctuations in the baryon-to-total mass ratio on cluster scales
(\cite{ev97}). Furthermore, dynamical processes operating differentially on the
baryonic and dark matter do not appear able to cause variations at the implied
level (i.e. \cite{me94}).

In general, clusters are found based on their optical appearance and/or their
X-ray emission, which are both baryonic in origin. However, baryons are a small
component of the total cluster mass ($0.060 \pm 0.003 h^{-3/2}$,
\cite{ev97}). Therefore, clusters may be very biased locations; measurements
made in these regions may not be representative of the whole universe. Are
clusters special places where there just happens to be large amounts of
baryonic matter?  Are there massive objects containing little or no baryonic
matter which we have yet to detect?  Is MS1224 an example of such a dark
object?  If so, this would represent a significant challenge for theories of
large-scale structure formation and might also imply that estimates of
$\Omega_0$ based on cluster masses are underestimated.

It is, therefore, important to confirm the weak lensing mass measurement of
MS1224+2007 with an independent study. In this paper we present a weak lensing
study of MS1224+2007 carried out with the MDM 2.4m. The observations and
reductions are described in \S \ref{observations} and \S \ref{profit}. The
shear and mass measurements are described in \S \ref{shear} and \S \ref{mass}
along with comparisons with previous lensing and virial mass
estimates. Conclusions and future work are presented in \S \ref{conclusion}

\section{Observations} \label{observations}

The cluster MS1224+2007 was observed using with Michigan-Dartmouth-MIT (MDM)
2.4m telescope on 8-9 Feb. 1998. The total exposure time was 21600s in R ($41
\times 900$s) and 4500s in B ($5 \times 900$s). The telescope was dithered
between exposures.  The ``Echelle'' 2048$^2$ thinned SITe CCD was used with
0.275\arcsec\ pixels. The seeing on the combined R image is 0.9\arcsec\ FWHM
and 1.4\arcsec\ for the combined B image.  The RMS sky noise values are 28.0 B
mag per square arcsecond and 28.1 R mag per square arcsecond. The two nights
were not perfectly photometric, the sky transparency varied by 9\% as
determined from monitoring 50 stars in the field of MS1224+2007 (due to
occasional small clouds passing through the field of view). We scale each image
to match the flux of the highest throughput images.

Despite the variations in sky transparency, 50 standard stars (\cite{la92})
were measured over the two nights. Transformation functions with linear color
and airmass terms are fit to the data resulting in an RMS of 0.03 mag and 0.013
mag for the B and R data respectively while the color terms are 0.035 and
0.013, respectively. Since the R-band data is much deeper than the B-band data
and the color terms are very small, we calibrate the data assuming a mean
galaxy color of $\langle$B--R$\rangle$=1.6. This introduces errors of $\pm
0.05$ and $\pm0.02$ for the B and R photometry respectively.



The reddening in this field is E(B-V)=0.04 yielding A$_B = 0.17$ and A$_R =
0.11$ (\cite{sc98}).

\section{Faint Galaxy Photometry and Analysis} \label{profit}

The faint galaxy analysis was carried out using the analysis software ProFit
(developed by the author). This software, starting with the brightest
detections, fits an analytical model to each object using weighted, non-linear
least squares, and subtracts the light from the image. It then proceeds to
successively fainter objects. Once it has detected and subtracted all the
objects in an image it replaces each in turn and refits and resubtracts until
convergence is achieved. The software outputs brightness, orientation,
ellipticity and other image parameters based on the fitted function.

\section{Gravitational Shear} \label{shear}

\subsection{Theory} \label{theory}

For gravitational lensing, the relationship between the tangential shear,
$\gamma_T$, and surface mass density, $\Sigma$, is (\cite{mi91}, \cite{mi95}),

\begin{equation}\label{escude}
\gamma_T(r) = \overline{\kappa}(\le r) - \overline{\kappa}(r),
\end{equation}

\noindent
where $\kappa = \Sigma/\Sigma_{crit}$, the ratio of the surface density to the
critical surface density for multiple lensing, and $r$ is the angular distance
from a given point in the mass distribution. The critical density depends on
the redshift distribution of the background galaxies. The first term on the
right is the mean density interior to $r$ and the second term is the mean
density at $r$. Therefore, the presence of a foreground mass distribution will
distort the appearance of background galaxies.  For a given coordinate on the
image ($\vec{r}$), the distortion quantity for the i$^{th}$ galaxy is:

\begin{equation} \label{disteqn}
D_i(\vec{r}) = {1-(b_i/a_i)^2 \over 1 + (b_i/a_i)^2} \times
{[\cos(2\theta_i)(\Delta{x_i}^2 - \Delta{y_i}^2) +
2\sin(2\theta_i)\Delta{x_i}\Delta{y_i}] \over \Delta{x_i}^2 + \Delta{y_i}^2},
\end{equation}

\noindent where $(b_i/a_i)$ and $\theta_i$ are the galaxy axis ratio and
position angle, respectively. $\Delta x$ and $\Delta y$ are the horizontal and
vertical angular distances from $\vec{r}$ to galaxy $i$. $D$ is related to the
tangential shear by (\cite{se95}):

\begin{equation}
<D(r)> = 2{\gamma_T(r)[1-\kappa(r)] \over [1-\kappa(r)]^2 + \gamma_T^2(r)}
\end{equation}

\noindent In the weak lensing regime $\kappa << 1$, and $\gamma_T << 1$,
$\gamma_T \approx <D>/2$. 

Before proceeding to a discussion of the shear measurments we discuss sources
of systematic errors.

\subsection{Point-Spread Function Anisotropy} \label{anisotropy}

\begin{figure}
\plotone{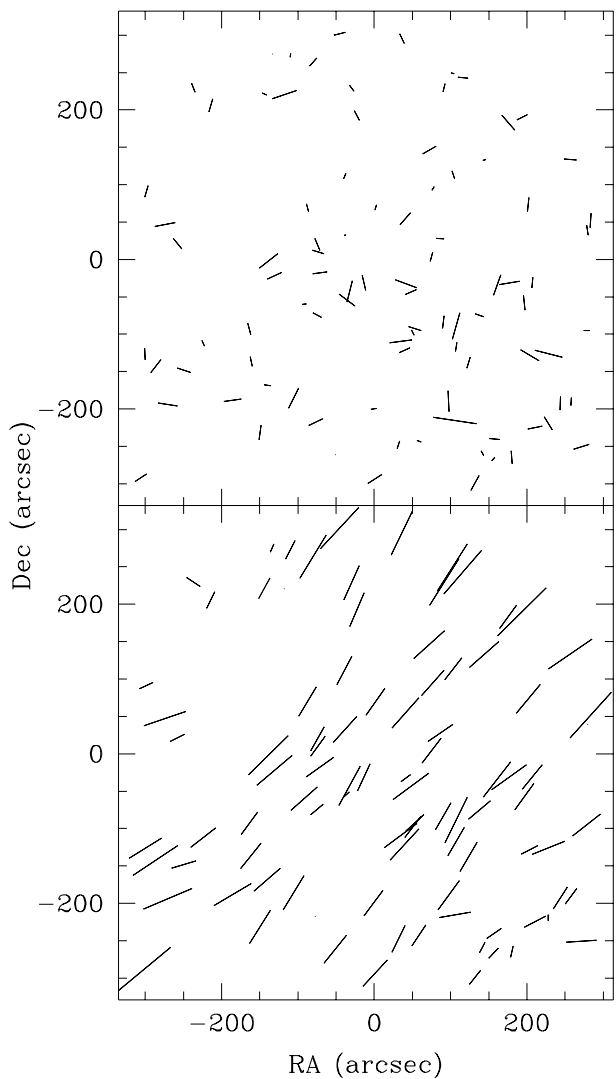}
\caption{Ellipticity and orientation for 87 stars in the combined R-band
image. The lower panel is the original image and the upper panel is after
correction (see text).  The maximum ellipticity is $\epsilon = 0.07$ and the
medians are $<\epsilon> = 0.035$ and $0.013$, for the original and corrected,
respectively. North is up and East is to the left.
\label{psf1}}
\end{figure}

In Figure \ref{psf1} the ellipticity and orientation for 87 stars with $16.4
\le$ R $\le 22.3$ in the combined R-band image are shown. There is a strong
anisotropy present in the point spread function (PSF) which, if not corrected
for, could bias estimates of the shear due to the cluster.  We correct for the
PSF anisotropy using the method outlined in \cite{fi97}, which involves
deriving a position dependent kernel which, after convolution with the image,
yields round PSFs. Figure \ref{psf1} shows the ellipticity and orientation for
the same stars after the correction. We discuss the effects of PSF anisotropy
on the shear and cluster mass estimates in \S \ref{measurements} and \S
\ref{2d}

\subsection{Seeing and Shear Polarizability} \label{seeing}

In order to measure the cluster mass using weak gravitational lensing we must
measure the shapes of background galaxies to estimate the shear (see \S
\ref{theory}). Even perfect galaxy measurements will result in biased shear
estimates for two reasons. The first is that the galaxy ellipticities will be
underestimated due to blurring by atmospheric seeing. The second is that the
response of a galaxy to a shear will depend on its ellipticity and orientation
with respect to the shear and hence the mean response will depend on the
intrinsic ellipticity distribution of the galaxies. In order to calibrate these
effects we carry out simulations using the F606W Hubble Deep Field Data (HDF)
(\cite{wi96}), using the techniques described in \cite{ka95}. This involves
stretching the HDF data by $1+\delta$, convolving with the PSF and adding
noise. The values of $D_i$ (see Equation \ref{disteqn}) are measured for each
galaxy and compared to the unsheared values of $D_i$. The quantity of interest
is the recovery factor, $C = \delta/<\Delta{D_i}>$ which is $<C> = 2.76 \pm
0.2$ for the present data (galaxies with $22.25 \le $R$ \le 24.0$ are used for
the shear analysis, see below). This correction does not take into account the
presence of stars which will further dilute the signal, however, this will be
minor for the magnitude range considered. 

\subsection{Cluster Galaxies} \label{clusgal}

Including cluster galaxies in the shear measurment will reduce the measured
value of the shear (assuming cluster galaxies are randomly aligned) by an
amount equal to the contamination fraction. This effect is likely to be a
function of radius since we expect the cluster galaxies to be centrally
concentrated within the cluster. For the shear analysis in this paper we use
galaxies in the range $22.25 \le $R$ \le 24.0$.  In order to get an estimate of
the number of cluster galaxies in this brightness range we assume that the
contamination is zero at the edge of our image. We then simply measure the
density of galaxies in our sample as a function of radius and fit a straight
line. This analysis reveals that at a radius of 27.5\arcsec\ about 22\% of the
galaxies in our field sample are probably cluster galaxies falling to zero (by
construction) at 275\arcsec. We adopt this correction for the rest of the
paper. If there are a significant number of cluster galaxies at the edge of the
image then this correction will be too small and we will be {\it
underestimating} the shear due to the cluster.

\subsection{Measurements} \label{measurements}

The value of the mean distortion, $<D(\vec{r})>$, for $\vec{r}$ equal to the
position of the central dominant galaxy (CDG) in MS1224+2007 for 874 galaxies
in the combined R image having $22.25 \le$ R$\le 24.0$ in the radial range
$27.5\arcsec\ \le r \le 275\arcsec$ is $<D_{raw}> = 0.034 \pm 0.006$; the
signal-to-noise is 5.5.  After correction for blurring by the PSF and shear
polarizability (see \S \ref{seeing}) the value is $<D_{cor}> = 0.094 \pm 0.017$
For comparison, the value of $<D(\vec{r})>$ for $\vec{r}$ at the CDG location
for the 87 PSF stars in the corrected image is $0.0009 \pm 0.002$, consistent
with zero and less than 3\% of $<D_{raw}>$. Given that the PSF-induced
distortion on the resolved galaxies is actually smaller than this, the residual
PSF anisotropy has little effect on our cluster mass estimate.

\begin{figure}
\plotone{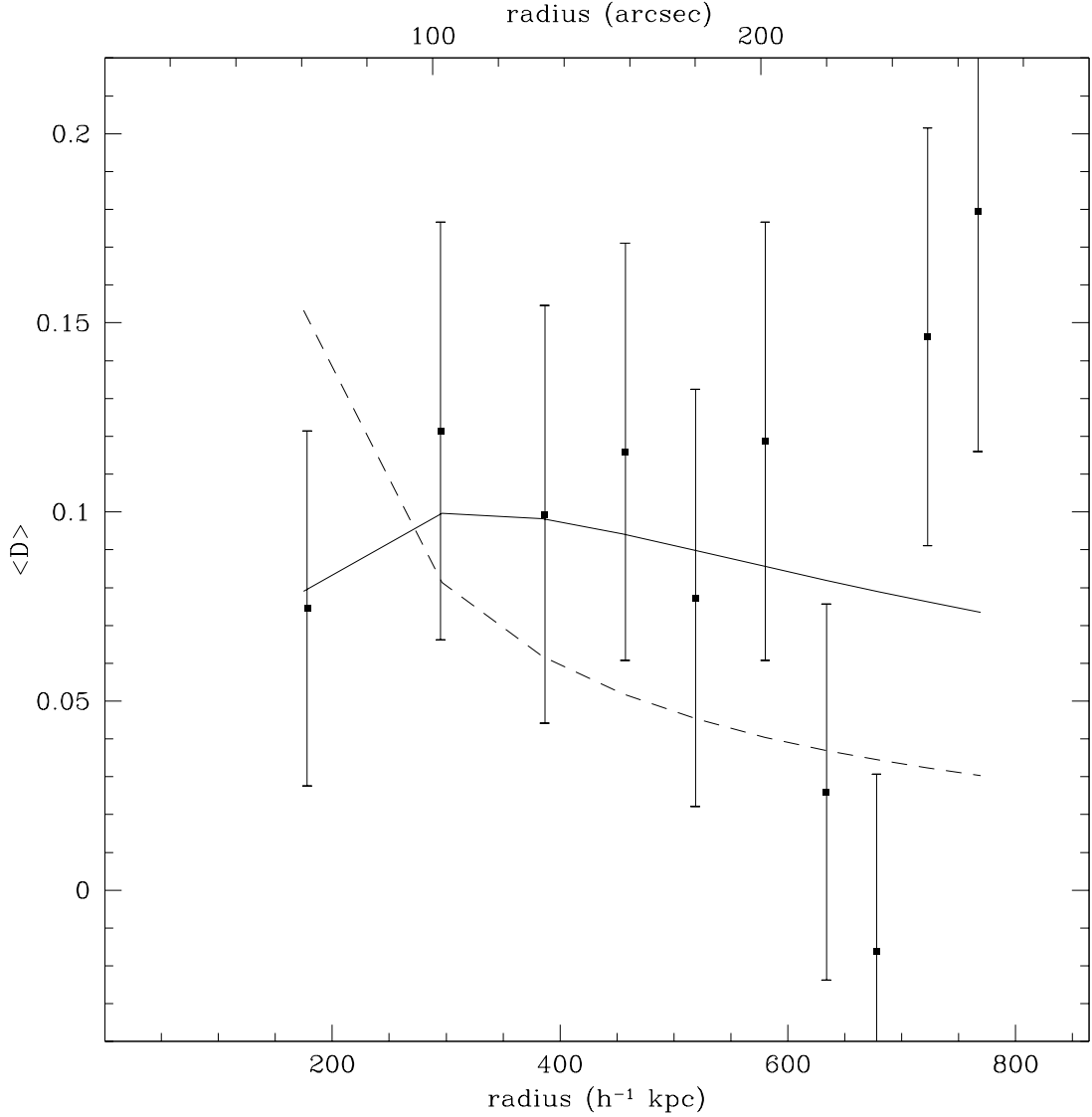}
\caption{A plot of the distortion, $<D>$, vs projected radius in radial bins
containing 87 galaxies each (outermost bin has 91) with $27.5\arcsec\ \le r \le
275\arcsec$ and $22.25 \le R \le 24.0$, centered on the dominant cluster
galaxy. The points are from the data ($1\sigma$ error bars) and the lines are
the expected distortion for two mass models. The dashed line is a SPIS while
the solid line is a NPIS with a core radius of $r_c = 45\arcsec=135 h^{-1}$
kpc. A recovery factor of $C = 2.76$ (see text) has been applied along with a
correction for contamination by cluster galaxies.
\label{fshear}}
\end{figure}

Figure \ref{fshear} shows $<D(\vec{r})>$ (corrected, for seeing, shear
polarizability and contamination by cluster galaxies) vs. projected radius for
$\vec{r}$ at the CDG position. The shear profile is very flat, the best fit
singular pseudo-isothermal sphere (SPIS) shown in Figure \ref{fshear} is ruled
out at the 95\% confidence level ($\chi^2 = 17.07$, nine degrees of
freedom). If we restrict ourselves to nonsingular pseudo-isothermal spheres
(NPIS) of form:

\begin{eqnarray}
\Sigma(R) & = & {\Sigma_0 \over [1 + (R/Rc)^2]^{1/2}},
\end{eqnarray}

then the best fit model has a core $R_c = 45^{+25}_{-15}\arcsec=135^{+65}_{-45}
h^{-1}$ kpc ($\chi^2 = 10.43$, nine degrees of freedom), and a central surface
density $\kappa(0) = 0.30 \pm 0.055$. The 95\% confidence limits on $r_c$ are
60 - 400 $h^{-1}$ kpc.

In order to convert the density profile into physical mass units we must know
the value of $\Sigma_{crit}$ which requires knowledge of the redshift
distribution of the background galaxies. There is no current redshift survey
complete to R = 24.0 (actually R = 23.9 because of reddening). We can get an
upper limit on $\Sigma_{crit}$ by using the redshifts from the Canada France
Redshift Survey (CFRS) (\cite{li95}). We derive R-band magnitudes by simple
averaging of their V and I photometry which should be accurate to 0.1-0.2 mag
(\cite{fu95}). Integrating $\Sigma_{crit}$ over the redshift distribution for
$22.25 \le$ R $\le 24.0$ yields $\langle\Sigma_{crit}\rangle = 1.50 h$
($\Omega_0 = 1.0$) and a mean redshift $\langle z \rangle = 0.69$. The value of
$\Sigma_{crit}$ will be an overestimate since the CFRS survey is incomplete for
much of this magnitude range. We perform a similar test with the deeper but
smaller Hawaii Deep Field Redshift Survey (interpolating R = 1/3B + 2/3I) and
obtain $\langle\Sigma_{crit}\rangle = 1.49 h$ and a mean redshift $\langle z
\rangle = 0.69$, although once again this survey is not complete over the
entire magnitude range. Another approach which should be somewhat less
sensitive to incompleteness at the faint end is to simply calculate the value
of $\Sigma_{crit}$ for an object at z = 0.69 (which is the mean of both the
redshift surveys). This yields $\Sigma_{crit} = 1.35 h$.  Finally we can use
theoretical models of galaxy formation and evolution to extrapolate the known
redshift distribution to fainter magnitudes. Adopting the model of \cite{gr96}
we find $\langle z \rangle = 0.73$ and $\langle\Sigma_{crit}\rangle = 1.44 h$
for the relevant reddening-corrected magnitude range.  Given the numbers above
we adopt $\Sigma_{crit} = 1.44 h$ with an uncertainty of around 10\%.



\section{Cluster Mass} \label{mass}

\subsection{Mass reconstruction} \label{reconstruct}

In the weak lensing regime, where $\kappa << 1$, the formula for the surface
mass density is (\cite{ka93}):

\begin{equation} \label{kseqn}
\kappa(\vec{r}) = {1\over
\overline{n}\pi}\sum_{i=1}^{N}{W(\Delta{x},\Delta{y},s) D_i(\vec{r})\over
\Delta{x_i}^2 + \Delta{y_i}^2}.
\end{equation}

\noindent 
where $N$ is the number of galaxies and $\overline{n}$ is the number density of
galaxies. Eqn. \ref{kseqn} assumes that the galaxies are intrinsically (in the
absence of lensing) randomly aligned. $W$ is a smoothing kernel which is
required to prevent infinite formal error.  In this paper we use a smoothing
kernel of the form (\cite{se95}):

\begin{equation}
W(\Delta{x},\Delta{y},s)=1-\left(1+{\Delta{x}^2+\Delta{y}^2 \over
2s^2}\right)e^{-(\Delta{x}^2+\Delta{y}^2)/2s^2},
\end{equation}

\noindent
where `$s$' is referred to as the ``smoothing scale''. The variance in the
dimensionless surface mass density is:

\begin{equation}
\langle\kappa^2\rangle = {\langle\gamma^2\rangle \over 8\pi\overline{n}s^2}.
\end{equation}

\begin{figure}
\plotone{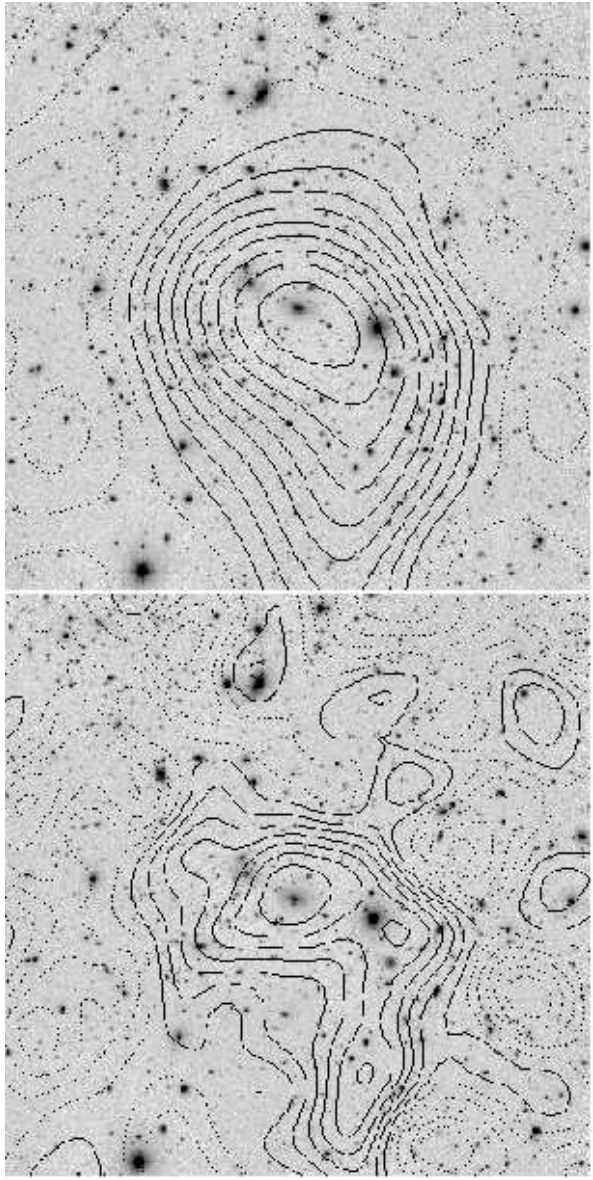}
\caption{Mass maps derived using Eqn. \protect\ref{kseqn} with smoothing scale
$s = 60\arcsec$ (top) and $s = 30\arcsec$ (bottom). A total of 1201 galaxies
with $22.25 \le$ R $\le 24.0$ are used in these reconstructions. The contours
are spaced in 1$\sigma$ intervals. The peak of the mass distribution is
consistent with the position of the central dominant galaxy. North is up and
East is to the left. The field is 9.6\arcmin\ on a side.
\label{massmap}}
\end{figure}

\noindent 2-d mass maps of MS1224+2007 are shown in Figure \ref{massmap} and
are discussed further in \S \ref{2d}.

\begin{figure}
\plotone{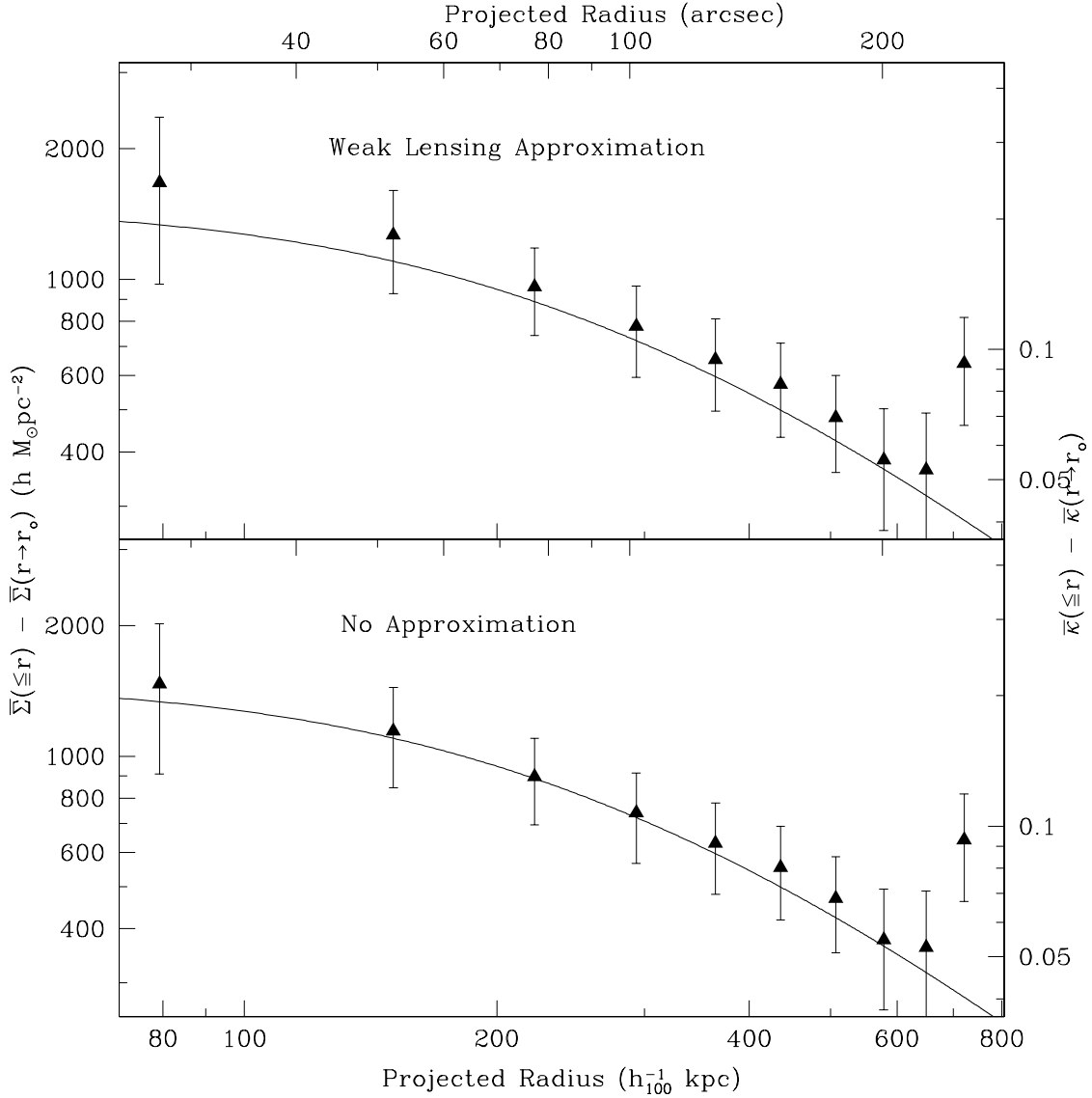}
\caption{The upper panel is the radial mass density profile (points) for
MS1224+2007 from Eqn. \protect\ref{denseqn} assuming $\kappa = 0$ ($r_{o} =
275\arcsec\ = 790 h^{-1}$ kpc). It is centered on the central dominant
galaxy. The points are the data for the cluster derived from 874 galaxies
having $22.25 \le$ R$\le 24.0$.  The lower panel is the radial density profile
from Eqn. \protect\ref{denseqn} using $\kappa$ derived from the best fit to the
shear shown in Fig. \protect\ref{fshear}. The value of $r_{o}$ is 275\arcsec.
A recovery factor of $C = 2.76$ (see text) has been applied. The solid lines
are the best NPIS model fits to the shear measurments shown in Figure
\protect\ref{fshear}.
\label{profilef}}
\end{figure}

Because of the smoothing kernel, plus biases introduced by edge effects in the
images, Eqn \ref{kseqn} is mainly useful for determining the 2-d shapes of mass
distributions. A less biased way of obtaining mass estimates as well as
azimuthally averaged density profiles is:

\begin{equation} \label{denseqn}
\overline\kappa(r \le r_i) - \overline\kappa(r_i \le r \le r_o) =
{r_o^2 \over N_{io}}\sum_{r_i \le r \le
r_o}{[1-\kappa(\vec{r})][1-\sqrt{1-D_i(\vec{r})^2}]
\over D_i(\vec{r})(\Delta{x_i}^2 + \Delta{y_i}^2)}
\end{equation}

\noindent
where $N_{io}$ is the number of galaxies between $r_i$ and $r_o$.
This is similar to the form employed by \cite{fa94} but is valid when $\kappa$
is not vanishingly small. Since $\kappa$ appears on the right hand side of the
equation, an iterative approach must be used to obtain the density profile.
Radial mass profiles for MS1224+2007 are shown in Figure \ref{profilef} and are
discussed further in \S \ref{profile}.

Galaxy distortion is insensitive to flat sheets of mass. Consequently, all mass
measurements described in this paper are uncertain by an unknown additive
constant. If there is a substantial flat component to the mass distribution our
mass estimates will be lower limits. 

\subsection{2-d Mass and Galaxy Maps} \label{2d}

The 2-d, KS mass map for MS1224+2007 is shown in Figure \ref{massmap}
superposed on the R-band image of the field.  This reconstruction uses 1201
galaxies with $22.25 \le$ R$ \le 24.0$. Before discussing the mass distribution
we check for potential biases introducecd by residual PSF anisotropies by
making a map with the 87 stars from the {\it corrected} image. The highest
value in this map is 8\% of peak value in the cluster map (equal to the
1.0$\sigma$ noise level of the mass map) and therefore all but the lowest
couple of contour levels should be unaffected by residual PSF anisotropy. For
comparison a map made from the same stars on the {\it uncorrected} image had a
peak value five times higher.

The mass distribution is peaked very near the position of the CDG and exhibits
an extension towards the south. In the higher resolution map this extension
appears as a distinct subcluster although its proximity to the edge of the
frame makes this conclusion uncertain. An additional subcluster appears to the
west of the CDG. This mass maps differs somewhat from the map of
\cite{fa94}. The peak of their mass distribution is centered east of the CDG
and appears to be extended to the North.

\begin{figure}
\plotone{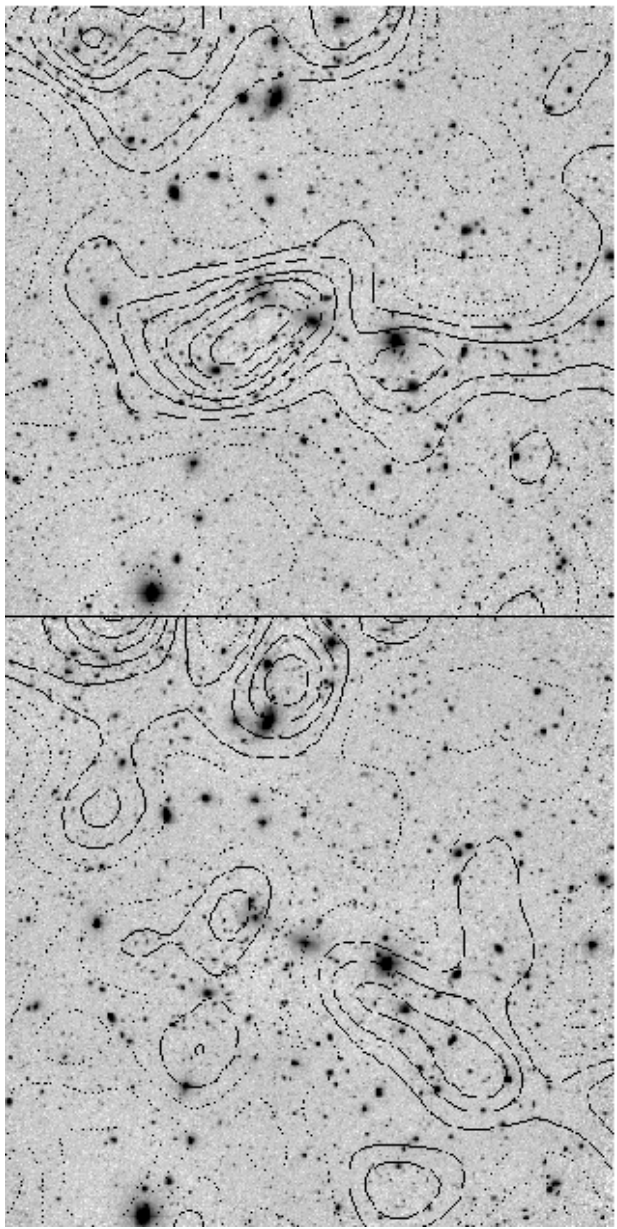}
\caption{Galaxy number density for galaxies in the range $17.28 \le$ R $\le
22.25$ (upper panel) and $22.25 \le$ R $\le 23.75$ (lower panel). Both have
been smoothed with a Gaussian having a scale of 30\arcsec. The contours are
spaced in 1$\sigma$ steps which corresponds to 1.2 and 1.3 per square arcmin
for the upper and lower panels, respectively \label{galdens}}
\end{figure}

\begin{figure}
\plotfiddle{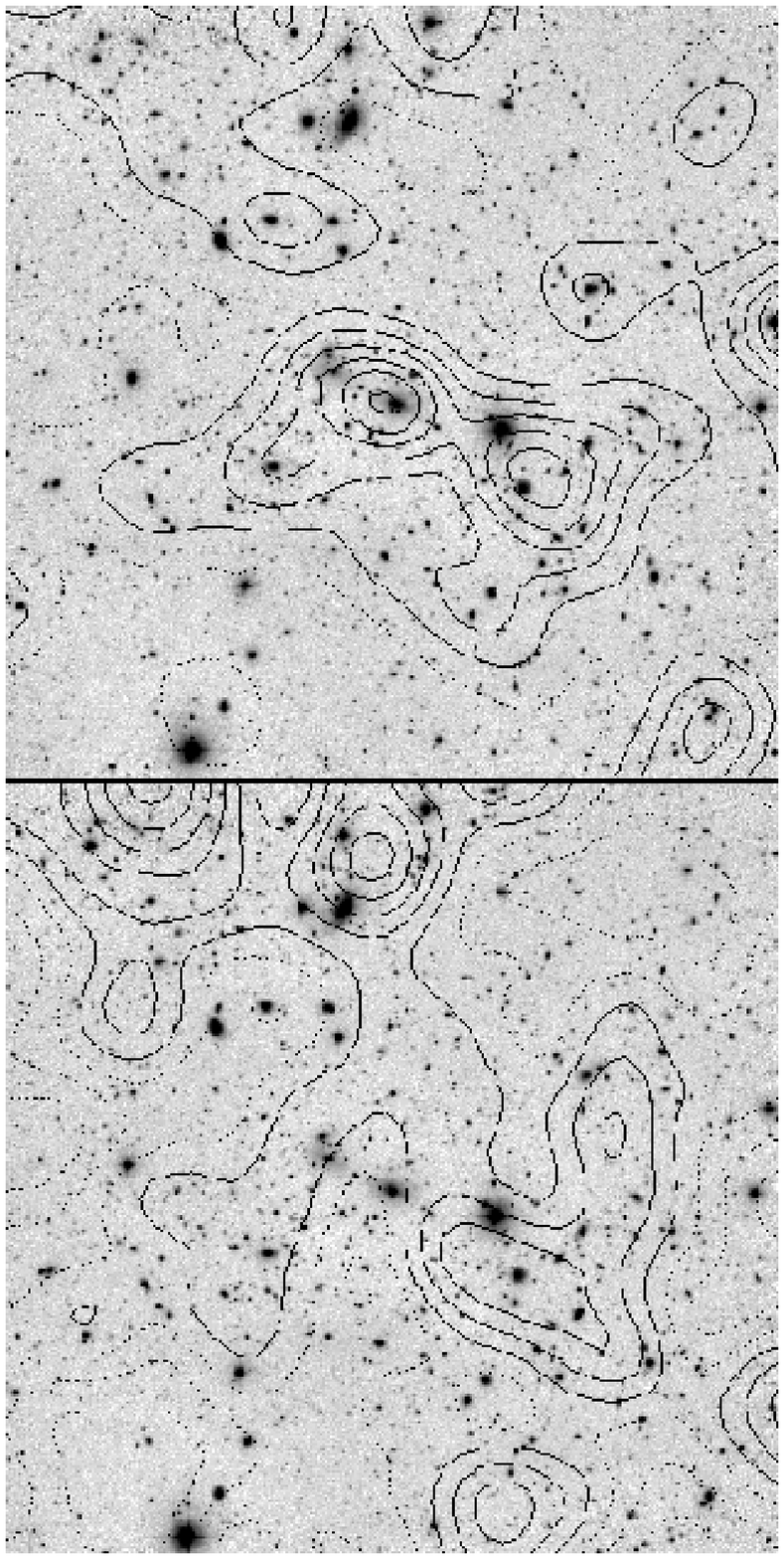}{7in}{0}{65}{65}{-170}{0}
\caption{Galaxy luminosity weighted number density for galaxies in the range
$17.28 \le$ R $\le 22.25$ (upper panel) and $22.25 \le$ R $\le 23.75$ (lower
panel). Both have been smoothed with a Gaussian having a scale of
30\arcsec. The contours are spaced in 1$\sigma$ steps which corresponds to 21.5
mag per square arcsecond (R-band reddening corrected) and 24.5 mag per square
arcsecond for the upper and lower panels, respectively \label{gflux}}
\end{figure}

In Figures \ref{galdens} and \ref{gflux} we show maps of the galaxy number
density and flux-weighted number density, respectively for two R-band magnitude
ranges (smoothed to match the high-resolution mass map). The bright number
density map is peaked slightly to the east of the CDG while the faint map seems
to have a concentration to the southwest of the CDG and a marginal signal in
the south. Both show evidence for significant galaxy clustering in the northern
portion of the field. The bright flux-weighted map appears to be bimodal with
two similar sized luminosity concentrations. One is centered very close to the
CDG (not surpising given the brightness of the CDG) and one is to the west. The
faint flux-weighted map looks a lot like the faint number density map with a
concentration to the southwest of the CDG and a significant feature to the
north and a less significant feature to the south.

It is difficult to draw strong conclusions from these maps without redshift
information. However, the bimodal central feature seen in both of the bright
galaxy maps and in the mass map seems to indicate that there is significant
subclustering within the cluster. Furthermore, there is strong evidence for a
large concentration of galaxies to the north. Because this concentration is
seen more prominently in the faint galaxies and is not seen (or marginally
seen) in the massmap it is likely that it is at a higher redshift than
$z=0.325$.

\subsection{Mass and Luminosity Profiles} \label{profile}

In the upper panel of Figure \ref{profilef} we show the azimuthally averaged
surface mass density profile centered on the CDG as derived from Equation
\ref{denseqn} using 874 galaxies having 22.25 $\le $R$ \le 24.0$.  We have
assumed that $\kappa = 0$ on the right hand side of the equation. The bottom
panel shows the density profile corrected using the best fit values of $\kappa$
from the fits of a NPIS to the shear (see \S \ref{shear}). The data have been
corrected for seeing, shear polarizability and contamination by cluster
galaxies.

\begin{figure}
\plotone{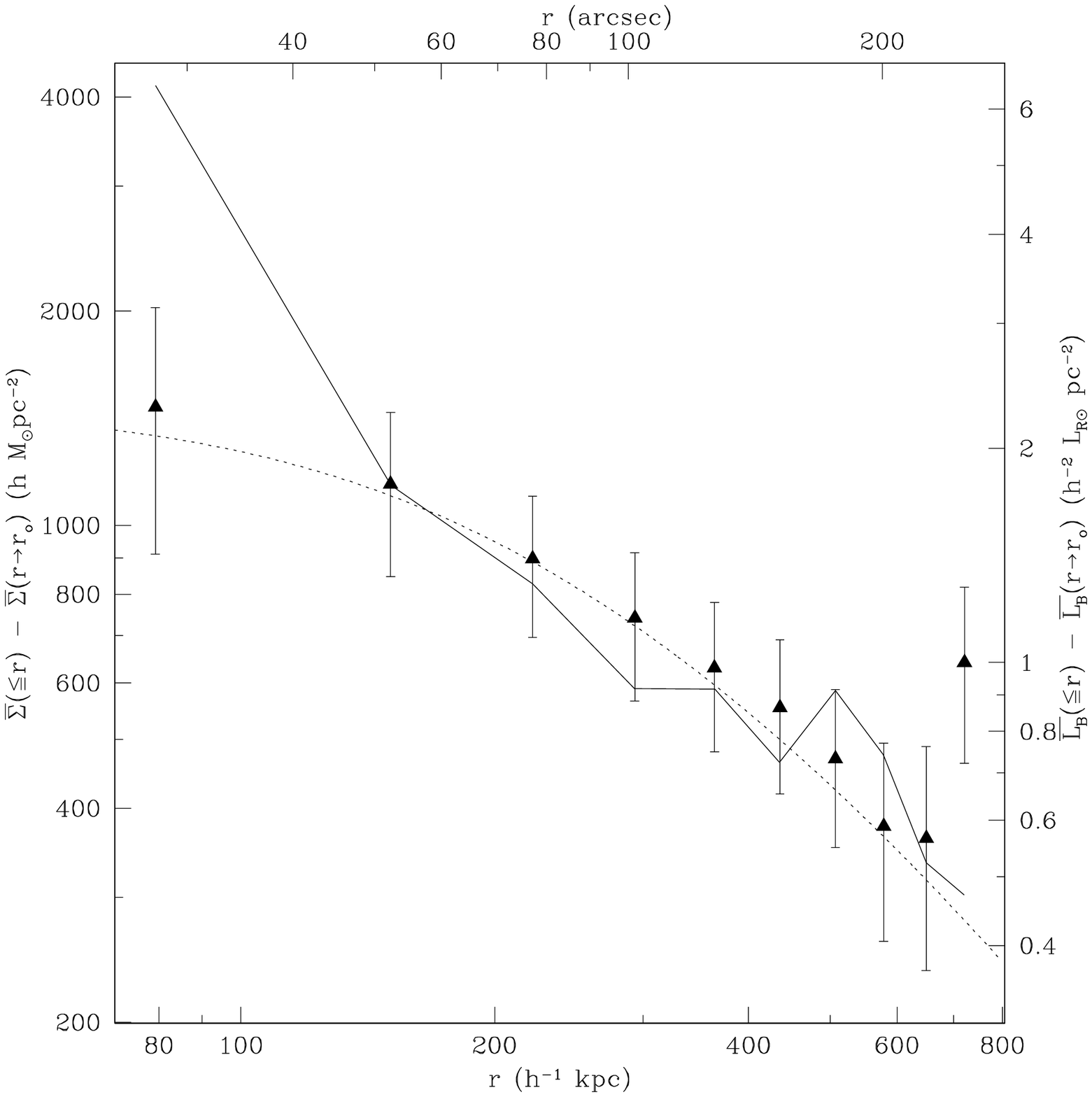}
\caption{Plot of projected cluster mass density (points) and projected
rest-frame R-band luminosity density (solid line) in galaxies. Both are plotted
as density contrasts.  The dotted line is the best pseudo-isothermal plus core
model fit to the shear measurments shown in Figure
\protect\ref{fshear}.\label{surf}}
\end{figure}

Figure \ref{surf} shows the mass density profile overplotted with the rest
frame R-band surface brightness profile. The surface brightness profile
includes light in galaxies with R $\ge 17.28$ (the magnitude of CDG) but does
not include possible diffuse light unassociated with galaxies. It is plotted as
a density contrast for comparison with the mass density profile. Aside from the
innermost regions where the light is completely dominated by the the CDG, it
appears that mass traces light.  We have applied a K correction to the R-band
photometry assuming the cluster light is dominated by early type galaxies of
K$_{cor} = 0.4$ (\cite{po97}) and an extinction correction of A$_R = 0.11$
(\cite{sc98}). The mass and light profiles are flatter than has been seen for
other clusters. Of course, subclustering will cause the azimuthally averaged
mass and light profiles to flatten out.

\begin{figure}
\plotone{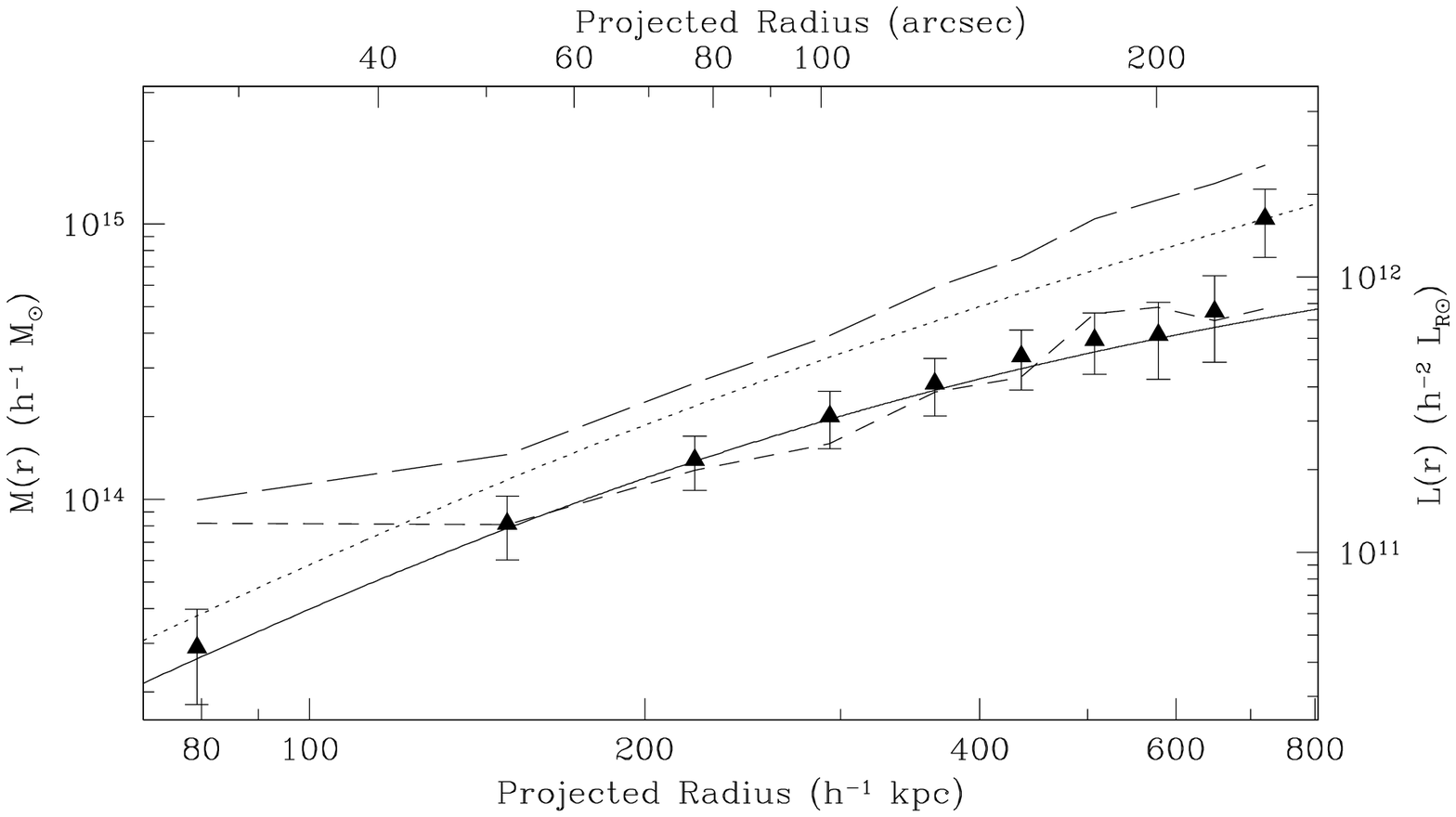}
\caption{Cumulative mass (points) and rest-frame R-band luminosity (short
dashes) profiles for MS1224+2007 based on $\pi r^2 [\bar{L}(\le r) -
\bar{L}(r\rightarrow r_{o})]$ and $\pi r^2 [\bar{\Sigma}(\le r) -
\bar{\Sigma}(r\rightarrow r_{o})]$, respectively (see Figure
\protect\ref{surf}). These are lower limits on the true cumulative
profiles. The solid line is the best-fit NPIS model. The dotted line shows the
true cumulative mass profile if the the NPIS model is a good description of the
data.  The long-dashed line is the cumulative luminosity profile assuming zero
background. This is an overestimate since field galaxies must have some
contribution to the R-band light.
\label{massf}}
\end{figure}

\begin{figure}
\plotone{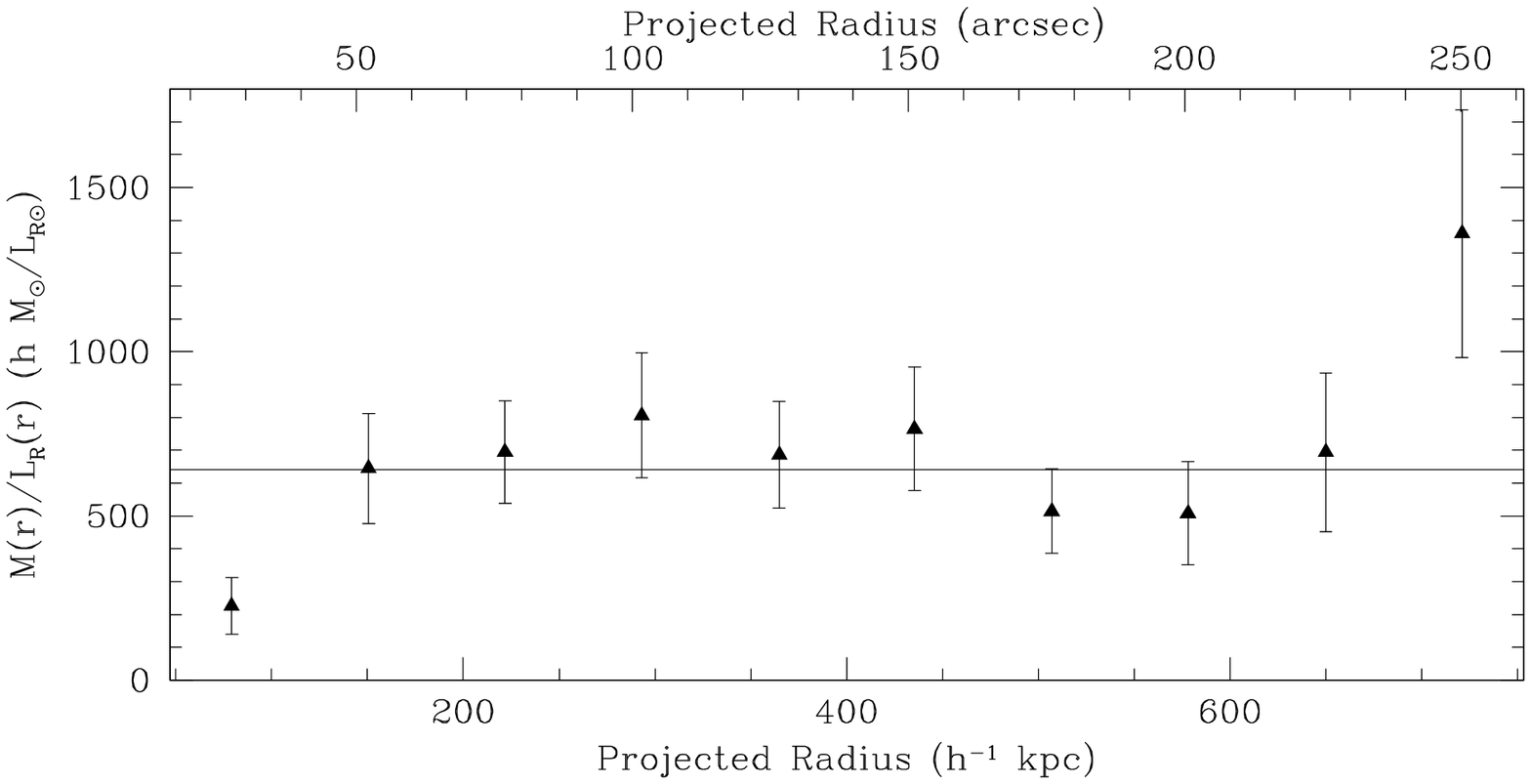} 
\caption{Cumulative mass-to-light ratio as a function of radius. Excluding
the innnermost and outermost points one gets M/L$_R = 640 \pm 150$
(M/L$_R)_\odot$. \label{ml}}
\end{figure}

Figure \ref{massf} shows the cumulative mass and luminosity profiles and Figure
\ref{ml} shows M(r)/L$_R$(r) as a function of radius. Within 100 $h^{-1}$ kpc
of the CDG the M/L is typical of cluster galaxies ($225 \pm 50$
(M/L$_R)_\odot$). However, beyond the region where the CDG light dominates, the
M/L increases rapidly and levels off. Ignoring the first and last points,
M/L$_R = 640 \pm 150$ (M/L$_R)_\odot$ (M$_{R\odot} = 4.31$) which is an
unusually high value. Assuming the cluster is dominated by early-type galaxies,
the evolutionary correction to $z=0$ is -0.363 mag (\cite{po97}) which yields
M/L$_R = 890 \pm 200$ (M/L$_R)_\odot$. The typical color of a nearby early-type
galaxy is (B-R) = 1.57 (\cite{fu95}) which gives an evolution-corrected M/L$_B
= 1300 \pm 300$ (M/L$_B)_\odot$ (M$_{B\odot} = 5.48$).

\subsection{Comparison with Previous Lensing Measurement}

In a previous study of MS1224+2007 (\cite{fa94}) a projected mass estimate is
given for $r \le 2.76\arcmin$ of M$(r \le 2.76\arcmin) = 3.5 \times 10^{14}
h^{-1}$ M$_\odot$ (no uncertainty is specified but it is at least 20\%). This
value is obtained without correction for the weak lensing approximation,
cluster galaxies, or mass in the control annulus which means it is an
underestimate. The mass we find with the same assumptions is M$=3.6 \pm 0.9
\times 10^{14} h^{-1}$ M$_\odot$, in excellent agreement with the previous
measurements. There is one caveat to this comparison in that different values
of $r_o$ (see Equation \ref{denseqn}) were used for the two estimates; $r_o =
400\arcsec$ was used in the previous study and $r_o = 275\arcsec$ is used in
this study. If we correct our value to $r_o = 400\arcsec$ by extrapolating the
NPIS model it increases the mass by approximately 20\%, still within 1$\sigma$
of the previous value. If we correct the raw value for the weak lensing
approximation it reduces the mass by only 3\%, and correcting for all the
previously mentioned systematics yields M$(r \le 2.76\arcmin) = 6.3 \pm 1.5
\times 10^{14} h^{-1}$ M$_\odot$.

We are also consistent with the value of M/L = 800 M$_\odot$/L$_\odot$ (no
bandpass specified) determined by \cite{fa94} even though the cluster
luminosity is calculated quite differently in this paper. In this paper the
cluster M/L was determined by measuring the mass and luminosities as density
contrasts centered on the CDG (see \S \ref{profile}). This means that we
include light emanating from clustered galaxies at redshifts other than that of
the main cluster provided that they are located near the cluster center (a
uniform background will be subtracted using this technique).  In \cite{fa94}
the luminosity estimate was based on 30 spectroscopically confirmed cluster
members from \cite{ca94} out of a sample of 75 measured redshifts to r=22.0 in
a $7\arcmin \times 9\arcmin$ field centered on the CDG.



\subsection{Comparison With Virial Mass Estimate} \label{virial}

In addition to the previous lensing mass estimate, there is a virial mass
estimate based on redshifts for thirty galaxies with radii $0\arcsec\ \le
275\arcsec$. The velocity dispersion is $802 \pm 90$ km s$^{-1}$
(\cite{ca96}). The velocity dispersion implied by our mass profile is dependent
on the nature of the phase space distribution function (DF) and the extent of
the cluster. If we assume an isotropic DF and an infinite cluster then the
implied dispersion is about 1380 km $s^{-1}$ averaged over the positions of the
galaxies with measured redshifts implying a mass over three times higher than
the virial mass estimate. If we truncate the mass profile at 3 $h^{-1}$ Mpc
then the projected velocity dispersion falls to 1290 km s$^{-1}$, still much
higher than the measured value.

There is evidence from both the mass map and galaxy maps for subclustering
within the cluster at z=0.325 and possibly at other redshifts along the
line-of-site. The redshift study of \cite{ca94} identified two groups of
galaxies at $z=0.22$ and $z=0.412$ for which they have estimated velocity
dispersions of 500 km s$^{-1}$ and 400 km s$^{-1}$, respectively. Subclustering
within the cluster complicates the virial analysis while clustering along the
line-of-site complicates the lensing analysis. For example, if the main cluster
actually consists of two large subclusters falling together in the transverse
direction then the virial analysis will quite likely underestimate the mass. If
there are groups in the foreground or background then the lensing mass for the
$z=0.32$ cluster will be overestimated, although the effects on the
mass-to-light estimates will depend on the redshifts of the groups relative to
the main cluster. Subclustering at the cluster redshift will {\it not} effect
the lensing M/L estimate.

The current number of measured redshifts is insufficient to attempt to
quantitatively disentangle the contributions from the various groups and
clusters. A full discussion will have to wait until we have completed a
photometric redshift survey of a large region centered on the cluster.




\section{Conclusions and Future Work} \label{conclusion}

In this paper we study the mass and light distributions in the $z=0.325$
cluster MS1224+2007 out to a projected radius of 800 $h^{-1}$ kpc by measuring
the gravitationally-induced distortions of background galaxies. We detect a
shear signal in the background galaxies in the radial range $27.5\arcsec\ \le r
\le 275\arcsec$ significant at the 5.5$\sigma$ level. The resultant mass map
(smoothed on 60\arcsec\ angular scales) exhibits an 8$\sigma$ peak centered on
the dominant cluster galaxy. A higher resolution (30\arcsec) mass map reveals
evidence for substructure which is even more strongly seen in the distribution
of galaxies. The lensing and redshift data combined indicate that there is
substructure at the cluster redshift and galaxy clustering in both the
foreground and background of the main cluster.

Assuming all the detected shear is due to mass at $z=0.325$ we find that,
except in the very central regions where the light from the CDG dominates, the
azimuthally averaged mass and light profiles follow one another with a
reddening and k-corrected mass-to-light ratio of M/L$_R = 640 \pm 150$
(M/L$_R)_\odot$.  The profiles are quite flat compared to other clusters,
disagreeing with a singular pseudo-isothermal sphere at the 95\% confidence
level. The best fit non-singular pseudo-isothermal sphere has a core radius of
$r_c = 135^{+65}_{-45} h^{-1}$ kpc. The flat profile is probably a consequence
of substructure.

Our mass and M/L estimates are consistent with the previous weak lensing result
of \cite{fa94}.


If we assume that the cluster is described by a an pseudo-isothermal sphere
with a non-singular core and has infinite extent then the velocity dispersion
implied by the lensing mass is almost 1400 km s$^{-1}$. Truncating the profile
at smaller radius reduces this dispersion; however, unless one truncates at
very small radius the lensing derived value remains much higher than the
measured dispersion of 802 km s$^{-1}$ (\cite{ca96}). A partial explanation
might be subclustering for which there is strong evidence. Infall of two or
more subclusters (approximately) perpendicular to the line-of-site could result
in the virial mass estimate substantially underestimating the cluster mass.

This cluster remains an anomalous object. It appears to be the highest M/L
cluster known and therefore is a candidate for a dark mass lacking baryonic
matter. This interpretation severely tests cosmological models which are unable
to produce such variations in baryonic fraction on cluster scales.  The
conclusion is weakened by a lack of information regarding foreground and
background clusters which would result in an overestimate of the cluster mass
and M/L.  The obvious next step in the study of this interesting and
controversial object would be to increase the number of galaxies with redshift
measurements.  This will allow us to look for galaxy clustering in redshift
space and disentangle the contributions of foreground and background mass
concentrations.  We can then derive definitive mass and mass-to-light estimates
for the primary cluster and whatever large mass concentrations exist along the
line-of-sight. The most efficient way to achieve these goals is with
photometric redshifts.

\acknowledgments

Support for this work was provided by NASA through grant \# HF-01069.01-94A
from the Space Telescope Science Institute, which is operated by the
Association of Universities for Research in Astronomy Inc., under NASA contract
NAS5-26555. Thanks to Caryl Gronwall for generating theoretical redshift
distributions. I acknowledge informative conversations with Gary Bernstein and
John Arabadjis.

{}

\clearpage

\end{document}